\documentclass[pdflatex,sn-basic,iicol]{sn-jnl}

\usepackage{graphicx}
\usepackage{supertabular}
\jyear{2022}%

\DeclareRobustCommand{\ion}[2]{%
\relax\ifmmode
\ifx\testbx\f@series
{\mathbf{#1\,\mathsc{#2}}}\else
{\mathrm{#1\,\mathsc{#2}}}\fi
\else\textup{#1\,{\mdseries\textsc{#2}}}%
\fi}
\newcommand{\vsini}{$v \sin i$}
\newcommand{\kms}{km\,s$^{-1}$}

\newcommand{\teff}{$T_{\rm eff}$}
\newcommand{\logg}{$\log g$}

\begin{document}

\title[UV Magnetometry]{Ultraviolet Spectropolarimetry: Investigating stellar magnetic field diagnostics}


\author*[]{\fnm{C. P.} \sur{Folsom}$^{1}$}\email{colin.folsom@ut.ee}

\author[]{\fnm{R.} \sur{Ignace}$^{2}$}

\author[]{\fnm{C.} \sur{Erba}$^{2}$}

\author[]{\fnm{R.} \sur{Casini}$^{3}$}

\author[]{\fnm{T.} \sur{del Pino Alem\'an}$^{4,5}$}

\author[]{\fnm{K.} \sur{Gayley}$^{6}$}

\author[]{\fnm{K.} \sur{Hobbs}$^{7}$}

\author[]{\fnm{R.} \sur{Manso Sainz}$^{8}$}

\author[]{\fnm{C.} \sur{Neiner}$^{9}$}

\author[]{\fnm{V.} \sur{Petit}$^{10}$}

\author[]{\fnm{M. E.} \sur{Shultz}$^{10}$}

\author[]{\fnm{G. A.} \sur{Wade}$^{11,7}$}

\abstract{
Magnetic fields are important for stellar photospheres and magnetospheres, influencing photospheric physics and sculpting stellar winds.  Observations of stellar magnetic fields are typically made in the visible, although infrared observations are becoming common.  Here we consider the possibility of directly detecting magnetic fields at ultraviolet (UV) wavelengths using high resolution spectropolarimetry, specifically considering the capabilities of the proposed Polstar mission.  UV observations are particularly advantageous for studying wind resonance lines not available in the visible, but they can also provide many photospheric lines in hot stars. 
Detecting photospheric magnetic fields using the Zeeman effect and Least Squares Deconvolution is potentially more effective in the UV due to the much higher density of strong lines. 
We investigate detecting magnetic fields in the magnetosphere of a star using the Zeeman effect in wind lines, and find that this could be detectable at high S/N in an O or B star with a strong magnetic field. 
We consider detecting magnetic fields using the Hanle effect in linear polarization, which is complementary to the Zeeman effect, and could be more sensitive in photospheric lines of rapid rotators. The Hanle effect can also be used to infer circumstellar magnetism in winds.  Detecting the Hanle effect requires UV observations, and a multi-line approach is key for inferring magnetic field properties. 
This demonstrates that high resolution spectropolarimetry in the UV, and the proposed Polstar mission, has the potential to greatly expand our ability to detect and characterize magnetic fields in and around hot stars.
}
%abstract length should be 150-250 words for ApSS

\keywords{
Ultraviolet astronomy (1736); Ultraviolet spectroscopy (2284); Spectropolarimetry (1973); Stellar magnetic fields (1610); Stellar winds (1636); Early-type stars (430); Instruments: Polstar; UV spectropolarimetry; NASA: MIDEX
}

\maketitle

%%%%%%%%%%%%%%%%%%%%%%%%%%%%%%%%%%%%%%%%%%%%%%%%%%
%%%%%%%%%%%%%%%%% BODY OF PAPER %%%%%%%%%%%%%%%%%%
%%%%%%%%%%%%%%%%%%%%%%%%%%%%%%%%%%%%%%%%%%%%%%%%%%

%%%%%%%%%%%%%%%%% INTRODUCTION %%%%%%%%%%%%%%%%%%

\section{Introduction}\label{Introduction}

Magnetic fields play an important role in stars across the HR diagram.  In massive O and B-type stars strong magnetic fields, when present, interact with strong stellar winds, modifying the structure of the wind, the angular momentum loss of the star, and even the mass-loss rate of the star (see \citealt{udDoulaTC} for a review). In intermediate mass stars, magnetic fields play a critical role in modifying diffusion and surface abundances, and creating Ap and Bp stars.  In lower mass stars magnetic fields are crucial for generating stellar coronae and winds, and controlling the angular momentum evolution of these stars. In very low mass stars, magnetic fields and the stellar activity they generate may also be important for the habitability of close-in exoplanets.  

The direct detection of magnetic fields relies on spectroscopy and, except for very strong magnetic fields (where Zeeman splitting is larger than all other line broadening sources), requires spectropolarimetry.  These magnetic fields are usually diagnosed through the Zeeman effect, although the Hanle effect is a valuable tool for solar observations and it could be useful for other stars provided sufficient spectropolarimetric data.  Existing spectropolarimetric observations have mostly been acquired in the visible wavelength range (e.g. with the ESPaDOnS, Narval, FORS1/2, or HARPSpol instruments).  Recently some spectropolarimetric observations have been extended into the infrared (IR), with new instruments such as SPIRou \citep{Donati2020}, and instruments that are coming online \citep[e.g. CRIRES+][]{Lavail2021}.  Observations in the IR can benefit cool stars with more flux in this wavelength range, and take advantage of the wavelength dependence of the Zeeman effect.  However, hotter stars have few spectral lines and much less of their flux in the IR.  Hot stars do have some wind lines in the IR, which can provide useful magnetospheric diagnostics, but most magnetic field observations are better performed at much shorter wavelengths.

There are no high-resolution, large wavelength coverage, spectropolarimeters that operate significantly into the ultraviolet (UV).  However, there are significant potential advantages for hot stars in observing at these short wavelengths.  Most obviously one can observe closer to the peak of the flux distribution of the star, additionally the density of spectral lines of hot stars is higher in the UV than in the visible.  

Several space projects for UV spectropolarimetry are currently under development. {\em Pollux} on a LUVOIR-type flagship mission at NASA \citep{ferrari2019}, {\em Arago} proposed for an ESA M mission \citep{Pertenais2017}, or the {\em Polstar} NASA MIDEX mission project \citep{Scowen2021SPIE-Polstar, ScowenTC} could make such spectropolarimetric observations at high resolution.
This work focuses on the capabilities of the {\em Polstar} mission, and has been performed in that framework, however the results are applicable to other similar proposed UV high resolution spectropolarimetric missions.
The proposed {\em Polstar} mission would provide spectra covering approximately 122 to 213 nm at a resolution of $\sim$33000. The instrument would provide total intensity Stokes $I$ spectra together with polarized Stokes $Q$, $U$, and $V$ spectra.

For the study of stellar magnetic fields, the biggest advantage of observing in the UV is the presence of resonance lines and the strongest wind lines.  In hot stars, there are no detectable resonance lines in the visible, thus the UV provides wind diagnostics (including magnetic field information) not available in the visible.  The UV also provides some advantages for detecting photospheric magnetic fields for hot stars, although this is possible with visible spectra.  Observations that constrain both photospheric and circumstellar winds simultaneously offer further advantages.  We first discuss detecting photospheric magnetic fields using the Zeeman effect and compare optical and UV models for hot stars in Sect.\ \ref{sec:lsd_magnetometry}, since this is the most common type of magnetic analysis currently.  In Sect.\ \ref{sec:circumstellar_magnetometry} we investigate the possibility of detecting magnetic fields in stellar winds through the Zeeman effect, using models of the magnetosphere and wind lines.  In Sect.\ \ref{sec:hanle} we discuss the Hanle effect in unresolved photospheres and winds, which is only detectable in UV lines of hot stars.

\section{Photospheric magnetometry in the UV} \label{sec:lsd_magnetometry}

\begin{figure}[tb]
\centering
\includegraphics[width=1.0\linewidth]{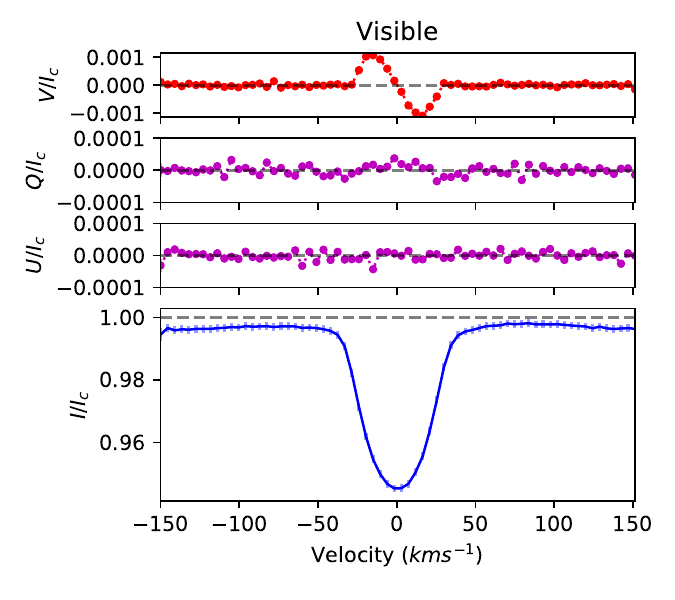}
\includegraphics[width=1.0\linewidth]{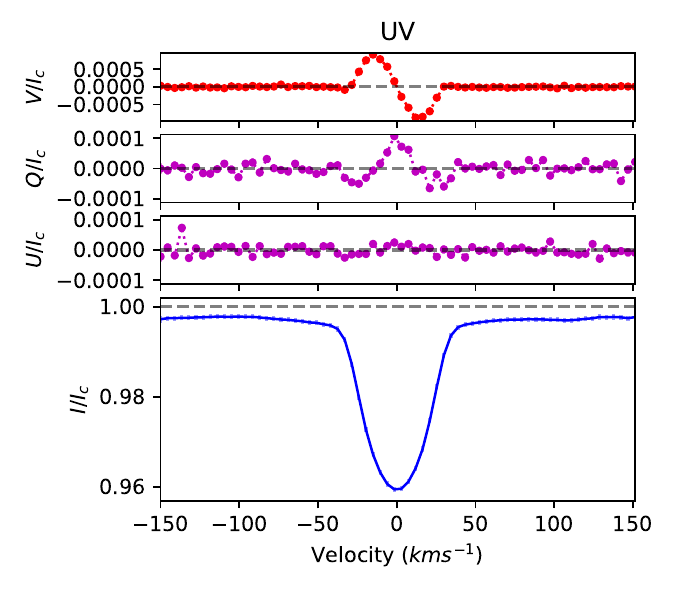}
\caption{LSD Stokes $I$, $Q$, $U$, and $V$ profiles, computed from simulated visible spectra (top) and UV spectra (bottom).  These models are computed for a B star with \teff~$= 20000$ K, \logg~$= 4$, \vsini~$= 30$ km/s, a polar field strength of $B_p = 3$ kG, and S/N~=~500 (at 160 and 500 nm).  Stokes $Q$ is definitely detected in the UV but only marginally detected in the visible model.  Stokes $U$ is undetected, which is expected as it is very weak for this particular magnetic geometry and orientation.
}
\label{LSD_IQUV_Bstar}
\end{figure}

\begin{figure*}[tb]
\centering
\includegraphics[width=1.0\linewidth]{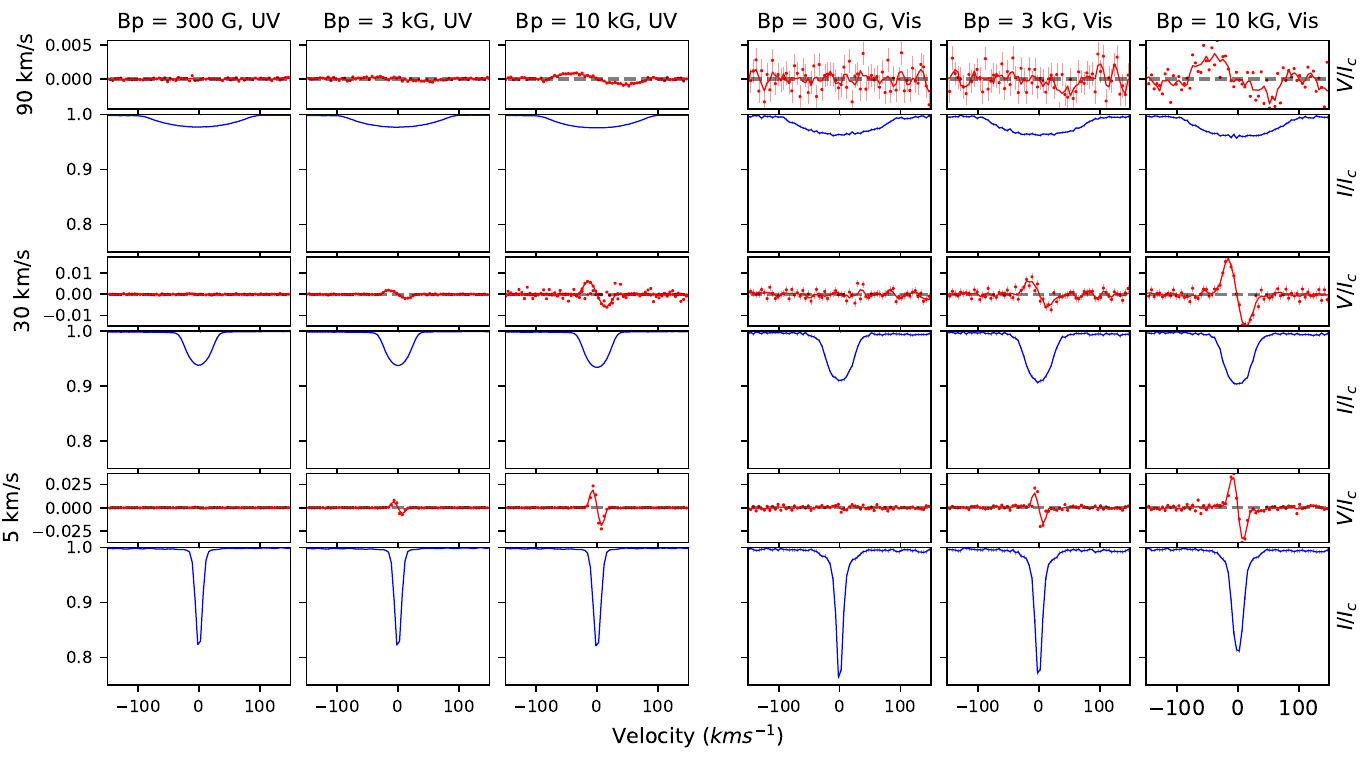}
\caption{LSD profiles for Stokes $I$ (blue) and $V$ (red), for a selection of different \vsini\ (vertically) and $B_p$ (horizontally) values.  Models for the UV range are on the left half and the visible range on the right half. All models are for \teff\ = 20000 K, \logg~$=4.0$, and a S/N of 100.}
\label{LSD_grid}
\end{figure*}

The Zeeman effect in spectral lines is routinely used to detect and measure magnetic fields at the surface of stars. If the magnetic field is very strong, and Zeeman splitting is larger than all other sources of line broadening (including \vsini), Zeeman split components of lines can be directly observed in the total intensity (Stokes $I$) spectra. If the field is weaker however, this splitting only appears as a small broadening of the line, which can be difficult or impossible to disentangle from other types of line broadening, since other broadening sources often dominate. However, the Zeeman components have different polarization properties (specifically the red and blue $\sigma$ components and the $\pi$ components, e.g.\ \citealt{2004ASSL..307.....L}). Therefore, measuring the polarization of spectral lines, i.e. measuring their Stokes parameters with spectropolarimetry, provides a very sensitive way to detect and characterize magnetic fields.  This approach can be highly effective, even when other sources of line broadening are larger than Zeeman splitting.

Least-Squares Deconvolution \citep[LSD,][]{1997MNRAS.291..658D, Kochukhov2010-LSD} is a powerful multi-line method for increasing the S/N of Zeeman signatures, allowing for magnetic field detections. 
It combines intensity or polarization signals from all available photospheric lines in a spectrum to produce a pseudo-average line profile and can be applied to all four Stokes parameters. 
This method models the spectrum as the convolution of a line profile with a series of delta functions, with wavelengths and amplitudes corresponding to individual lines (a `line mask'). It uses a linear least squares approach to fit the line profile to the observation, essentially deconvolving the line profile from the observation using the line mask.  The result is similar to a cross-correlation technique where the template is a series of delta functions.  Interpreting an LSD profile as a real line profile relies on some assumptions, most generally that all the lines used have the same shape, differing only in amplitude.  More specifically, this implies that the lines form in same regions of the atmosphere with the same magnetic fields, and that the magnetic field is weak enough that differences in Zeeman splitting patterns of different lines are negligible (i.e.\ that only differences in effective Land\'e factor matter). In practice LSD can still be useful when modest violations of those assumptions occur but using very different lines, such as photospheric lines and wind lines, or Balmer lines and iron lines, must be avoided.  LSD can be applied to spectra in all four Stokes parameters, although different line masks are needed to account for the wavelength and Land\'e factor dependence of the Zeeman effect.

LSD has been successfully applied to detect and characterize magnetic fields using Visible \citep[e.g.][]{1997MNRAS.291..658D, Wade2000-LSD-4Stokes} and more recently IR \citep[e.g.][]{Martioli2020-LSD-in-IR, Moutou2020-LSD-in-IR, Petit2021A&A-LSD-in-vis-IR} spectropolarimetry. However, it has not been applied in the UV yet, due to a lack of observations. 

The large number of photospheric lines available in the UV domain potentially makes LSD even more powerful in this waveband than in the visible. Figure~\ref{LSD_IQUV_Bstar}  compares model $I$, $Q$, $U$, and $V$ LSD profiles obtained for a B star in channel 1 of {\em Polstar} with a model in the visible (described further in Sect.\ \ref{sec:synthetic-tests-LSD}). For {\em Polstar} we used a resolution of R=33000 spanning 122-213 nm and for the visible we used R=65000 covering 370-990 nm, which corresponds to the resolution of ESPaDOnS. ESPaDOnS \citep{Donati2003-espadons} is one of the best high-resolution spectropolarimeter currently available for visible light and is installed at the Canada-France-Hawaii Telescope. 

\subsection{Wavelength scaling and S/N considerations}
\label{sec:analytic-LSD}

There are some general trends worth considering when assessing the practicality of observing photospheric magnetic fields in the UV.  
The photon energy shift from the Zeeman effect is wavelength independent, therefore the wavelength shift it induces is proportional to $\lambda^2$.
Most other line broadening processes (including thermal, turbulent, rotational, and usually instrumental broadening)
produce a fixed fractional change in wavelength, so their wavelength shifts are proportional to $\lambda$.
Thus in the weak magnetic field approximation, when the wavelength shift from Zeeman splitting is less than other local line broadening processes, the degree of circular polarization seen in Stokes $V/I$ profiles is proportional to $\lambda$. 

At first glance this would appear to penalize the UV domain, relative to the optical,
but the detectability of Zeeman splitting primarily depends on the S/N of the Zeeman-induced polarized photons, which is governed by photon counts not
degree of polarization alone.  Hence there are other contributing factors, such as stellar photon flux and the amount of Zeeman-sensitive
line opacity in the UV versus optical regimes. 

In general, stellar flux distributions are complex, however for hot stars, we can consider the Rayleigh–Jeans tail as an adequate approximation for purposes of comparing the UV to the optical.  Then the power per wavelength bin is proportional to $1/\lambda^4$.  For S/N considerations we are most interested in the rate of photons per detector pixel (or spectral resolution element), and converting to a rate of photons per wavelength bin causes this to become  $1/\lambda^3$.
The spectral resolution and pixel sizes are also proportional to $1/\lambda$ for most high-resolution spectrographs, including {\em Polstar}, thus the rate of photons per spectral pixel goes as $1/\lambda^2$.  Assuming a well calibrated detector with a sensitivity that is independent of wavelength, and that noise is dominated by shot noise (i.e. the noise, and S/N, goes as $\sqrt{N}$), the noise per spectral pixel is therefore proportional to $1/\lambda$.

By comparison, the Zeeman signal for a line of given fractional line depth (which is how lines are binned in the LSD approach used here) depends on the excess photon count of one circular polarization over the other, within some part of the line.  As mentioned above, that
fractional excess is proportional to $\lambda$.  
But the total line photon count is proportional to the continuum photon flux per wavelength bin, 
again proportional to $1/\lambda^3$, times the linewidth, proportional to $\lambda$.
Combining these shows that the signal in excess photon counts in one polarization is proportional to $1/\lambda$, just as was true for the noise.
Hence the S/N for Zeeman detection in hot stars using lines of similar fractional depth favors neither the UV nor the optical.

Now, for practical purposes, it is certainly
easier to build a large optical telescope, which can be ground based, than a UV telescope, which requires a space mission.  
Thus there will be large differences in collecting area.  For example ESPaDOnS, one of the most efficient and widely-used high-resolution spectropolarimeters, is on the 3.6 m Canada-France-Hawaii telescope.  The proposed {\em Polstar} mission would have a 0.6 m mirror, so
36 times smaller collecting area.  This would produce a daunting limitation for using the UV to observe photospheric magnetic fields,
if the same number of
lines of the same fractional depths were available in all spectral domains.

However, in hot stars the number of spectral lines available is much larger in the UV, and the lines are stronger.  Photospheric magnetic analyses rely on multi-line techniques, except for rare cases of extremely strong magnetic fields or exceptionally bright stars.  The most popular is
the LSD method, which we use here.
We extracted line lists from the Vienna Atomic Line Database \citep[VALD,][]{Ryabchikova2015-VALD3} using `extract stellar' requests, as the basis for LSD line masks.  Lines were then removed if they were blended with telluric bands, or exceptionally broad photospheric features (e.g.\ Balmer lines, C {\sc iv} and Si {\sc iv} resonance lines in the UV).  Only lines with a VALD line depth parameter $>0.1$ were retained.  For a model with \teff~$= 20000$ K, \logg~$=4.0$, and solar abundances this produces 7722 lines in the UV range of {\em Polstar}, but only 220 lines in the visible range of ESPaDOnS.  Using a larger depth cutoff in the UV of 0.3 would still leave more than a factor of 10 difference in the number of available lines. 

\begin{figure}[tb]
    \centering
    \includegraphics[width=1.0\linewidth]{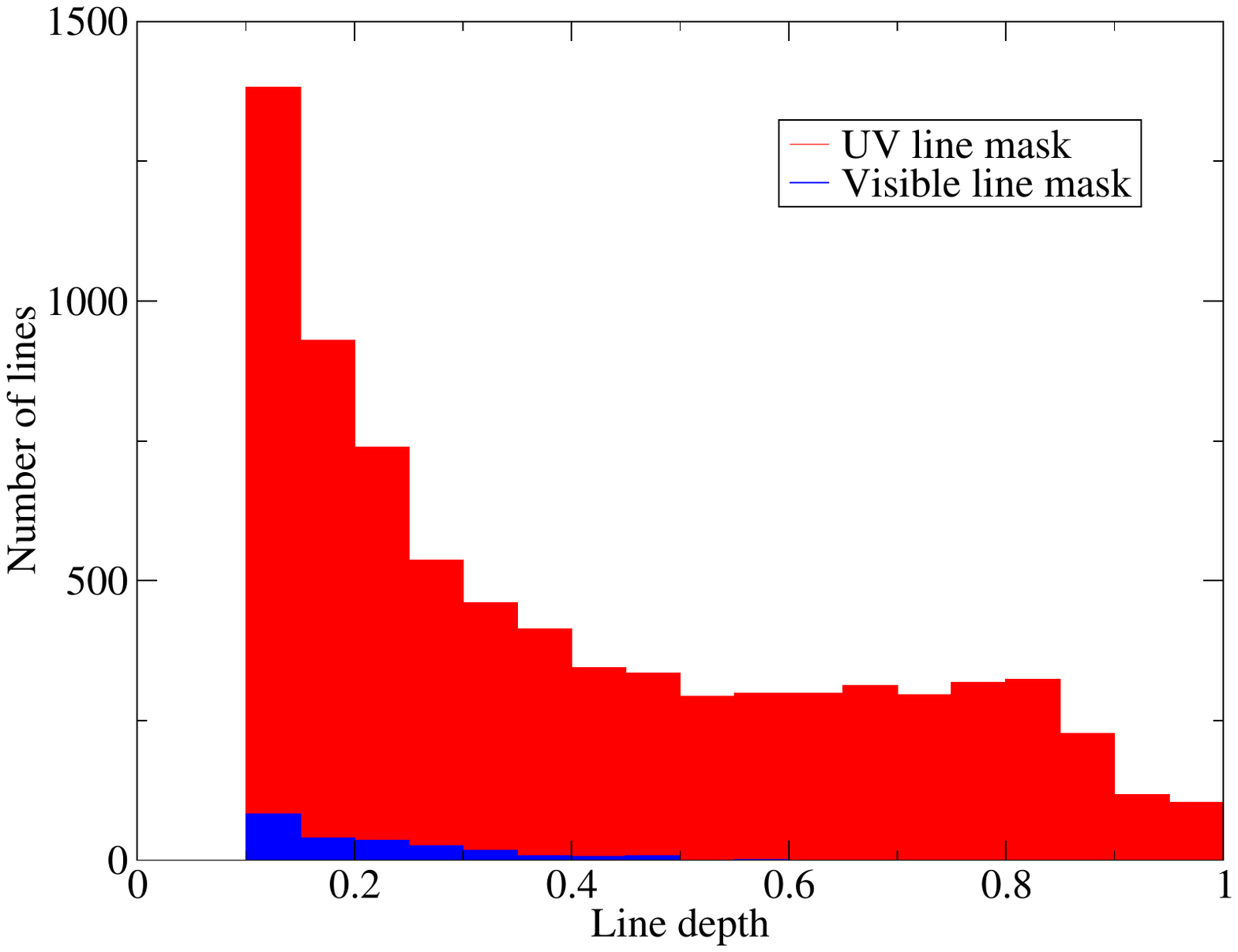}
    \caption{Distributions of line depths for UV and visible a line masks, using a model with \teff~$= 20000$ K, \logg~$=4.0$, and solar abundances.  }
    \label{fig:line-depths}
\end{figure}

In addition to having more lines available in the UV, the lines are stronger.  Histograms of the line depth parameter for the UV and visible line masks are compared in Fig.~\ref{fig:line-depths}.  This line depth parameter is calculated before most line broadening processes are included, thus rotational, turbulent, and instrumental broadening will reduce line depths in observed spectra, but by the same amount in both spectral
regimes.  
In order to quantify the impact of the larger line depths, we calculate 
$N^{-1/2}\,\sum_{i=1}^{N} d_{i} g_{i}$, where the sum of the line depth ($d_i$) times the effective Land\'e factor ($g_i$) is proportional to the total available signal, divided by the square root of the number of lines used as proportional to the noise (the wavelength dependence of the signal and noise compensate for each other, as discussed).  Comparing this quantity from the UV to the visible we find a factor of 10.8 increase in the ratio of signal photons to the shot noise in the UV.  

That S/N boost exceeds the factor 6 hit stemming from 
the factor $\sim 36$ reduction in collecting area, making the UV an effective regime for photospheric Zeeman studies.
This significant increase in line opacity in the UV provides many additional benefits for studying the circumstellar environment
of hot stars, so photospheric Zeeman studies are not the primary goal of UV spectropolarimeters like {\em Polstar}, but nevertheless
such studies justify the plan to use {\em Polstar's} high-resolution channel 1 in some of its observations of magnetic stars.

\subsection{Synthetic tests}
\label{sec:synthetic-tests-LSD}

These analytic considerations demonstrate that LSD should be effective in the UV, however they rely on several approximations.  
As a more detailed test of the efficiency of LSD in the UV, we apply LSD to model spectra calculated with the {\sc Zeeman} spectrum synthesis code \citep{Landstreet1988-Zeeman, Wade2001-Zeeman}.  
{\sc Zeeman} performs polarized radiative transfer including the Zeeman effect, and produces photospheric spectra assuming local thermodynamic equilibrium (LTE).  For these calculations atomic line data from the Vienna Atomic Line Database \citep[VALD,][]{Ryabchikova2015-VALD3} was used, along with model atmospheres from ATLAS9 \citep{Kurucz1993-ATLAS9-SYNTHE}.  
The limitation of LTE models may be important for particularly hot models (\teff\ $> 30000$ K). However, the LSD process for such hot stars avoids emission lines, wind lines, and particularly strong lines with large wings.  Thus places where wind and non-LTE effects are the most important are not used for LSD and our magnetic analysis, and the overall impact of non-LTE effects is reduced.

A simple magnetic geometry was used, consisting of a dipole field inclined at $\beta=90^\circ$ to the stellar rotation axis, the stellar rotation axis inclined at $i=90^\circ$ to the line of sight, and a rotation phase with the positive magnetic pole oriented towards the observer.  The rotation axis is along the positive Stokes $Q$ direction, which puts the rotation equator is along the negative $Q$ direction.  Most hot star magnetic fields are dominantly dipolar with smaller contributions from higher multipoles \citep[e.g.][]{Kochukhov2002-ZDI-alpha2CVn, 2015MNRAS.451.2015O, Grunhut2021-Plaskett}, so the adopted geometry is somewhat simplified but still realistically representative.  Three magnetic field strengths at the pole ($B_p$) were used: 300 G, 3 kG, and 10 kG.

A small grid of models was calculated for different \teff\ values of 10000, 20000, and 30000 K, all at \logg\ $=4.0$, and for \vsini\ of 5, 30, and 90 \kms.  In all cases solar abundances were assumed.  Many magnetic stars have peculiar abundances, but the specific abundances vary between stars, and in most cases produce stronger lines and hence more detectable magnetic fields.

\begin{figure*}[tb]
    \centering
    \includegraphics[width=1.0\linewidth]{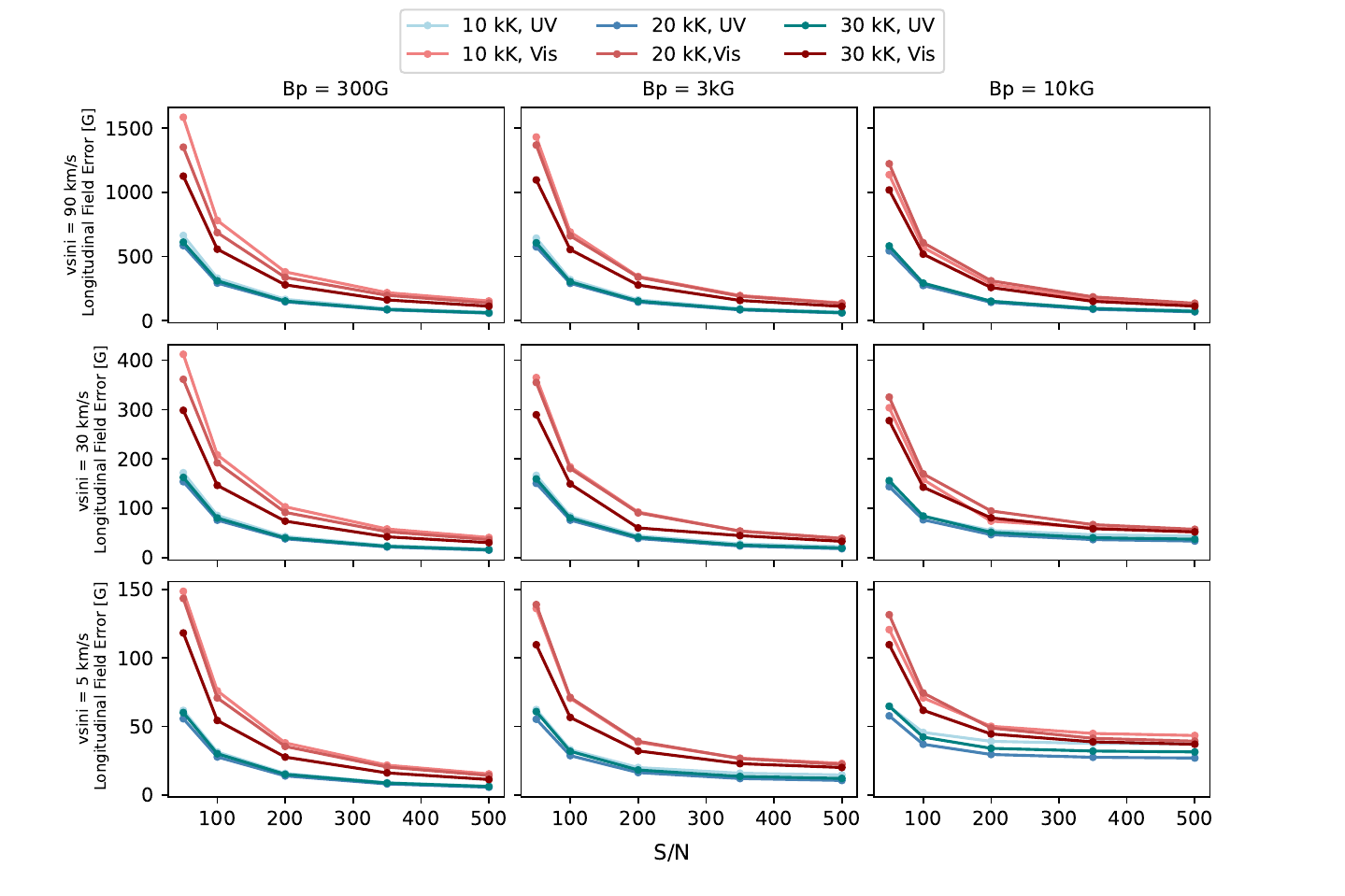}
    \caption{Uncertainties on longitudinal magnetic field values, calculated from LSD profiles for a range of model spectra. These provide a measure of the magnetic precision achievable.  Results for models at three different values of \vsini, $B_p$, and \teff\ are shown, for both UV and visible (Vis) spectra. 
}
     \label{fig:Bl-errors}
\end{figure*}

\begin{figure*}[tb]
    \centering
    \includegraphics[width=0.7\linewidth]{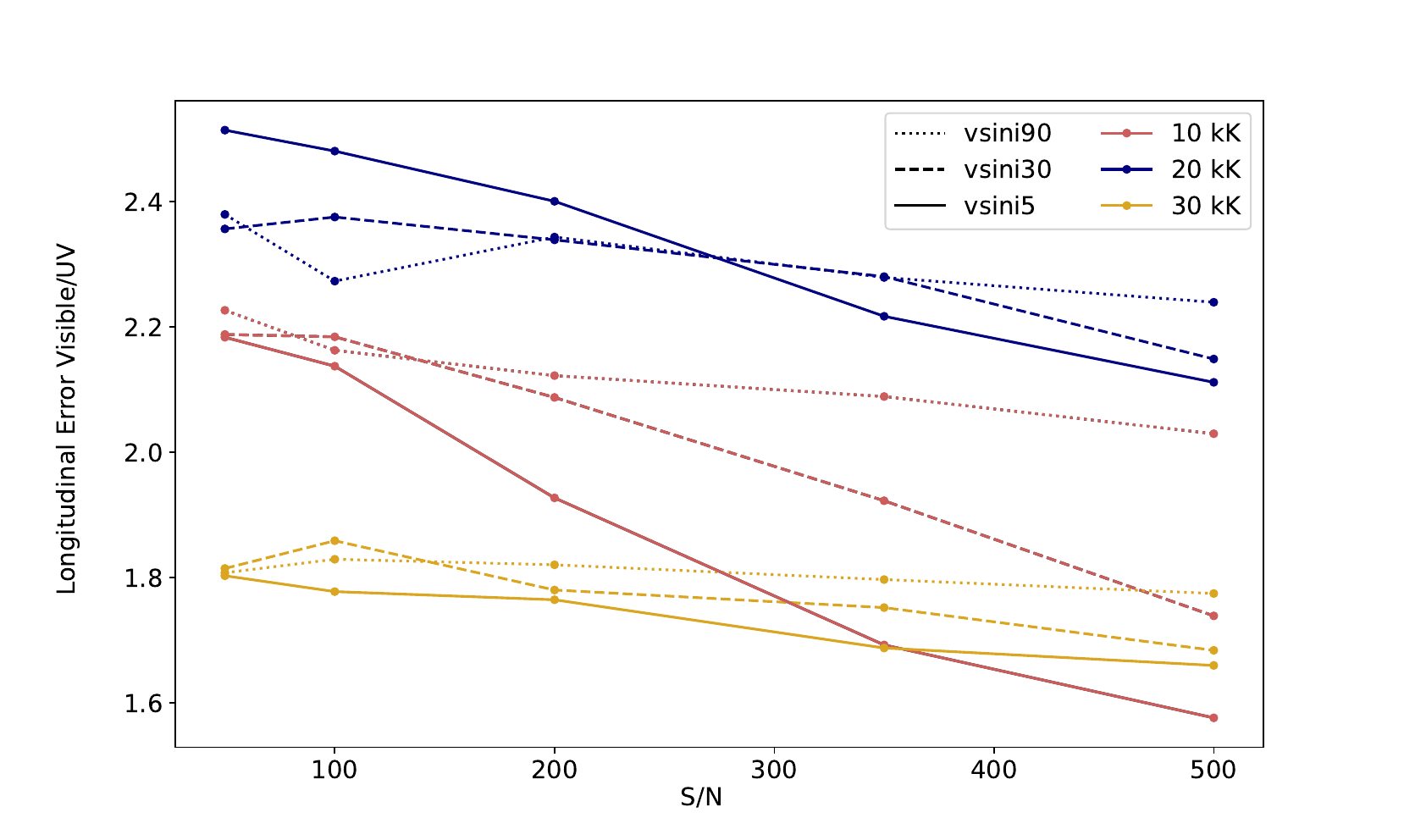}
    \caption{Ratios of the uncertainty on $B_\ell$ from the visible range to the corresponding value from the UV range.  This illustrates the improvement in magnetic precision that could be achieved with UV observations.  Models are shown for three different \vsini\ and \teff\ values for $B_p = 3$ kG ($B_p$ has a relatively small impact on the uncertainty reached).
}
     \label{fig:BlErr-ratios}
\end{figure*}

Synthetic spectra were calculated for the proposed {\em Polstar} channel 1 wavelength range (122-213 nm) and resolution (33000), and also for a representative ESPaDOnS resolution (65000) and wavelength range (370-670 nm; ESPaDOnS extends further to the red, but hot stars have very few lines and low S/N in that region).  The synthetic spectra were resampled to representative mean pixels sizes for {\em Polstar} (4.5 \kms) and ESPaDOnS (1.8 \kms).
Synthetic noise was added as a function of the flux per detector pixel, for several realistic S/N levels, to produce model observations. The S/N was scaled as the square root of the flux per detector pixel (assuming detector pixel sizes are constant in velocity), however this does not account for the wavelength dependence of instrumental efficiency, or a wavelength dependence in interstellar extinction or atmospheric transmission. The S/N values reported here are those obtained at 160 and 500 nm for the UV and visible, respectively.  \citet[][]{ShultzTC} find S/N values in this range to be realistic for exposure times of up to one hour with {\em Polstar}, for a substantial number of stars, based on observations from the International Ultraviolet Explorer and model fluxes from Tlusty.

LSD was applied to the model observations, generating LSD profiles, and allowing us to assess the detectability of signals in Stokes $V$, as well as $Q$ and $U$.  Line masks were created using data from VALD, starting from an `extract stellar' request for the \teff\ and \logg\ of the corresponding model spectrum.  Only photospheric lines have been included in the line mask, and hence in the LSD profiles. Exceptionally broad features, and any lines blended with them, were removed (H lines, some He lines, some other resonance lines in the UV such as the \ion{C}{iv} doublet at 1548.187 \& 1550.772 \AA\ and \ion{Si}{iv} doublet at 1393.755 \& 1402.770 \AA). Lines in the visible that would be blended with strong telluric lines were also removed from the mask. 
Following \citet{Wade2000-LSD-4Stokes} the Stokes $Q$ and $U$ LSD profiles were calculated using a line weight of $d g^2 \lambda^2$, for a line depth $d$, wavelength $\lambda$, and the effective Land\'e factor $g$ (Stokes $V$ used the standard weight $d g \lambda$).  
Normalizing values of $d=0.4$ and $g=1.2$ were used with a normalizing $\lambda = 160$ nm in the UV and $\lambda=500$ nm in the visible.
Resulting LSD profiles in all four Stokes are shown in Fig.~\ref{LSD_IQUV_Bstar}, while Stokes $I$ and $V$ for a wider selection of the grid of models is shown in Fig.~\ref{LSD_grid}.

Longitudinal magnetic field values ($B_\ell$) were calculated from the LSD profiles using Eq.~1 of \citet{Wade2000-highPrecision-Bz}.  This represents the line of sight component of the magnetic field integrated over the stellar disk and weighted by local brightness.  The uncertainty on $B_\ell$ provides a useful measure of the magnetic precision achieved in an observation.  
The error bars on the LSD profile also provide a measure of the precision of a magnetic detection, however they depend on the normalization parameters adopted for the profile.  The $B_\ell$ values and uncertainties are independent of this normalization \citep{Kochukhov2010-LSD}, and can be more directly compared with theoretical magnetic field strengths.

The uncertainties we find on $B_\ell$ for the grid of models are presented in Fig.~\ref{fig:Bl-errors}, over a range of S/N values.  The best achievable uncertainty depends strongly on \vsini, since as \vsini\ increases a line is spread over more pixels and the amplitude of a signal is reduced.  The uncertainty is relatively independent of the magnetic field strength, except for very high values ($B_p = 10$ kG) at lower \vsini, where resolved Zeeman splitting becomes important, limiting the effectiveness of LSD.  In such a case, studying individual lines with a good S/N becomes both practical and more informative.  The uncertainty depends weakly on \teff\ for the range of values investigated here, with changes driven by the number of lines available for LSD and the relative strengths of those lines.  For even hotter values (e.g. 40000 K), the uncertainties would rise considerably for visible spectra, due to the decrease in the number of lines available.  Preliminary testing suggests that UV spectra would fare better, with the uncertainty rising by a much smaller factor, since there appear to be many more lines available for LSD.  A consistent trend is that UV spectra provide smaller uncertainties than visible spectra at the same S/N for all \teff, \vsini, and $B_p$ values considered.  This is illustrated in Fig.~\ref{fig:BlErr-ratios}, where the uncertainty on $B_\ell$ from the visible spectra divided by the uncertainty from the UV spectra is plotted, for spectra with the same model parameters.  This demonstrates that the uncertainties achieved using UV spectra are consistently a factor of $\sim$2 smaller than than from the visible.  

The Zeeman effect in spectral lines scales with wavelength, the amplitude of the polarimetric signal being proportional to wavelength for weaker magnetic fields.  This leads to lower amplitude Stokes $Q$, $U$, and $V$ signals in the UV than in the visible, by a factor of 2-4 depending on the wavelengths considered. However, for hot stars there are many more lines available in the UV, thus the gain provided by multi-line techniques like LSD outweighs the loss in the amplitude of the polarimetric signal.  This leads to an improved sensitivity to magnetic fields of hotter stars in the UV.
This is illustrated in Fig.~\ref{LSD_grid}, where the amplitude in Stokes $V$ is lower for the UV profiles than the visible, but the noise is decreased even further, leading to more significant detections in the UV. 
These results do not account for differences in photon count rate per spectral pixel between the visible and UV, since they are for comparable S/N in both wavelength ranges.  
In order to assess the efficiency of LSD due only to the available lines, not including other wavelength dependent effects,
and to compare more directly with Sect.~\ref{sec:analytic-LSD}, one can multiply the ordinate of Fig.~\ref{fig:BlErr-ratios} by a factor of $\sim$3 to remove the wavelength dependence of the Zeeman effect.  In that case we find the gain in magnetic sensitivity provided by LSD to be $\sim$5-7 higher in the UV, which suggests our estimate from analytic considerations in Sect.~\ref{sec:analytic-LSD} of 10.8 is overly optimistic.  However this is still sufficient to compensate for a difference in mirror diameter of 6, and the advantages of the UV likely become stronger at \teff~$> 30000$ K due to the dwindling number of lines in the visible.

Spectropolarimetric observations can used to derive the geometry and strength of the magnetic field on a stellar surface.  This requires a time series of observations covering the rotation period of the star.  
Simple dipolar models can be inferred from the variations in $B_\ell$, while more complex models directly fit a series of LSD profiles, in order to reconstruct the magnetic field using tomographic techniques. 
Zeeman Doppler Imaging \citep[ZDI; e.g.][]{Donati-Brown1997-ZDI, Piskunov2002-ZDItechnique, Kochukhov2002-ZDI-alpha2CVn, 2006MNRAS.370..629D} is the most frequently used method for deriving maps of a surface magnetic field, and it has been applied to a wide range of stars.  The spacial resolution of the map depends on the spectral resolution of the observations, with high (R~$\geq 30000$) resolution need for good spacial resolution.  UV instruments may be very useful for analyses like ZDI, given the improved effectiveness of LSD in the UV over the visible for hot stars, but UV instruments must have high resolution for this kind of imaging to be viable.

These results demonstrate the increased effectiveness of LSD in the UV compared to the visible for O, B, and A stars, even though the {\em Polstar} wavelength range is much shorter than that of ESPaDOnS, due to an increase in the number and strength of the available spectral lines. This leads to an improved ability to detect and precisely characterize the magnetic fields of hot stars.

%%%%%%%%%%%%%%%%%%%%%%%%%

\section{Circumstellar magnetic fields using the Zeeman effect in UV wind lines}\label{sec:circumstellar_magnetometry}

In massive stars, the shapes of the wind-sensitive resonance lines observable at UV wavelengths are sensitive to the kinematics of the stellar wind, and so provide a diagnostic of the density and velocity structure of the circumstellar plasma. 
Resonance lines, which provide the most sensitive wind diagnostics, are only available in the UV for hot stars.
When coupled with theoretical models, lines such as {C~\textsc{iv} $\lambda\lambda$154.8, 155.1} nm, {Si~\textsc{iv} $\lambda\lambda$139.3, 140.2} nm, and {N~\textsc{v} $\lambda\lambda$123.8, 124.2} nm can be used to produce quantitative estimates of key parameters like the stellar mass-loss rate and wind terminal velocity. 

In magnetic massive stars, the field confines and channels the stellar wind into a magnetosphere, with a significantly more complex density and velocity structure than that of a spherically symmetric analog. This produces distinct changes within the UV wind-sensitive line profiles that have been observed \citep[e.g.][]{mar13,naz15,2019MNRAS.483.2814D,dav21,2021MNRAS.506.2296E} and modeled \citep[e.g.][]{udDoula2013,Hennicker2018,erb21} with various numerical techniques. Recently, \citet{erb21} performed a detailed, systematic parameter study of the various factors that affect line profile formation in magnetic massive stars.  

The UV line synthesis technique reported by \citet[][the \textsc{uv-adm} code]{erb21} can be extended to model magnetospheric polarization using UV wind lines, following the method outlined by \citet{Gayley2010} and \citet{Gayley2017}. Although often challenging to detect, Zeeman splitting is present in the spectral lines of magnetic massive stars \citep[e.g.][and see Section~\ref{sec:lsd_magnetometry} above]{2009ARA&A..47..333D}. The split line components are circularly polarized, which is then detected and measured using Stokes~$V$ $= I_{L} - I_{R}$ profiles. \citet{Gayley2017}'s method calculates the antiderivative of the Stokes~$V$~polarization profile \citep[see also][]{Gayley2010,Gayley2015,Kochukhov2015}. The UV-ADM code is used to determine the field-weighted intensity along rays traversing the magnetospheric structure, and the synthetic Stokes~$V$~profile is then obtained from the derivative with respect to wavelength.

\begin{table}[tb]
\begin{center}
\caption{Adopted stellar parameters for synthetic Stokes~$V$ profiles in Figure~\ref{fig:stokesv_uvadm}. The quantity $R_{\rm A}$ is the Alfv\'en radius \citep[e.g.][]{udDoula2002} corresponding to each model.
\label{tab:modelparams_stokesv_uvadm}}
\begin{tabular}{lcc}
\hline \hline
& \textbf{B-type Star} & \textbf{O-type Star} \\
$\log(\dot{M})$ & -10 $M_{\odot}$~yr$^{-1}$ & -7 $M_{\odot}$~yr$^{-1}$ \\
$v_{\infty}$ & 1200~km~s$^{-1}$ & 2700~km~s$^{-1}$ \\
$R_{\ast}$ & 4~$R_{\odot}$ & 10~$R_{\odot}$ \\
$R_{\rm A}$ & 40~$R_{\ast}$ & 9.5~$R_{\ast}$ \\
\hline
\end{tabular}
\end{center}
\end{table}

\begin{figure*}[tb]
\centering
\includegraphics[width=1.5\columnwidth]{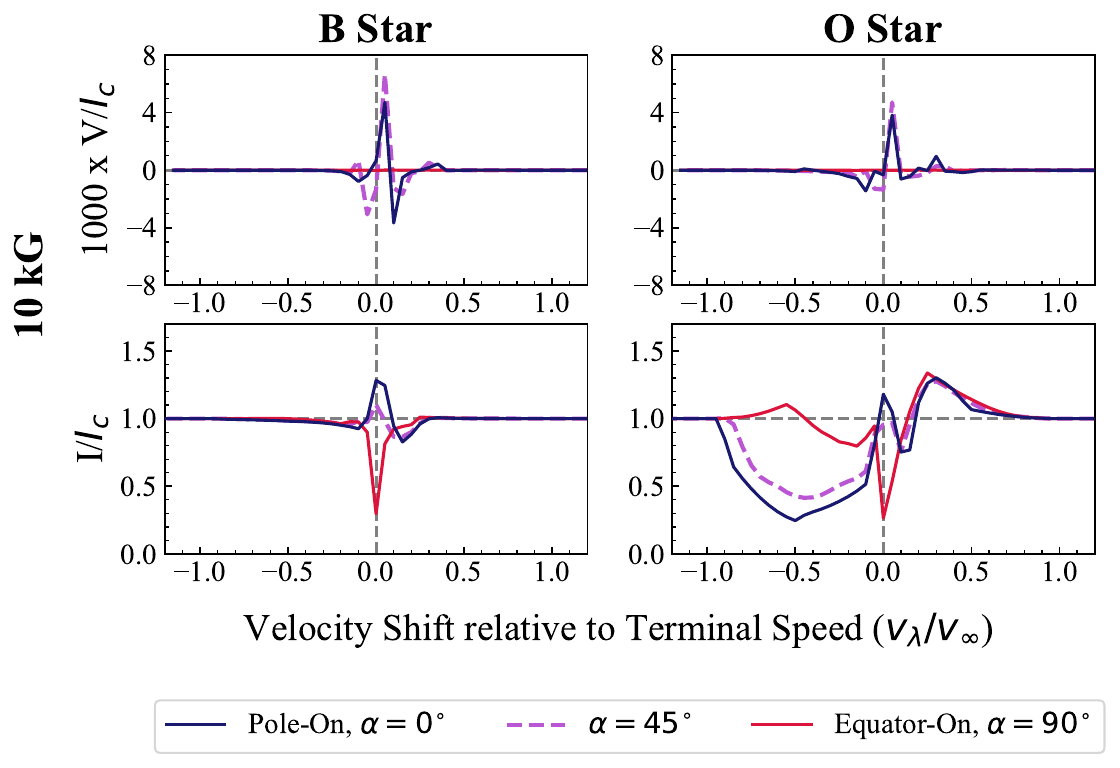}
\caption{
Synthetic Stokes~$V$ (top row) and intensity (bottom row) profiles created using the UV-ADM code for representative O and B-type stars with a surface field of $B_{\rm d} = 10$~kG. Profiles are provided for magnetospheric viewing angles of $\alpha$ = 0$^{\circ}$ (pole-on; blue solid lines), $\alpha$ = 45$^{\circ}$ (purple dashed lines), and $\alpha$ = 90$^{\circ}$ (equator-on; red solid lines). The Stokes~$V$~profiles use the red component of the wind-sensitive C~\textsc{iv}~doublet as a representative spectral line. The quantity $v_{\lambda}$ (wavelength expressed in velocity space) is plotted as a fraction of the terminal speed the wind would have in the absence of a magnetic field.}
\label{fig:stokesv_uvadm}
\end{figure*}

Figure \ref{fig:stokesv_uvadm} shows six Stokes $I$ profiles synthesized\footnotemark\ 
\footnotetext{Following \citet[][]{erb21}, the models reported here use the Analytic Dynamical Magnetosphere (ADM) formalism \citep{Owocki2016} to calculate the density and velocity structure of the magnetosphere. However, a snapshot of a 3D magnetohydrodynamic (MHD) simulation of the magnetosphere could also be used, which will be discussed in greater detail in a forthcoming paper.}
for the representative O- and B-type stars listed in Table \ref{tab:modelparams_stokesv_uvadm}. 
For these stars, we assume a surface field strength of $B_{\rm d}=10$~kG, and calculate line profiles for viewing angles\footnote{The viewing angle $\alpha$ is defined to be the angle between the line-of-sight to the observer and the north magnetic pole \citep{erb21}.} of $\alpha$ of 0$^{\circ}$ (pole-on; blue solid lines), 45$^{\circ}$ (purple dashed lines), and 90$^{\circ}$ (equator-on; red solid lines). The profiles were calculated using line strengths and Land\'e factors corresponding to those of the red component of the wind-sensitive C~\textsc{iv}~154.8, 155.1 nm doublet.

The Stokes~$V$ signature from the equator-on view of each sample magnetosphere is effectively null, as required by the top-bottom symmetry with reversed line-of-sight magnetic field. However, the amplitude of the Zeeman signature in the pole-on view for a 10~kG field suggests that detection of a Stokes~$V$ signature is reasonably attainable at an expected $V/I$ sensitivity level of about 0.1\%.

The current models only consider the polarization arising from resonance scattering in the wind (that is, they do not include a photospheric contribution), and they assume that the Zeeman shift does not affect the total photon scattering -- only the wavelengths at which scattering occurs -- in the circularly polarized Zeeman split components.
Hence, the total area under the Stokes $V$ profile should be zero, consistent with its representation as a derivative of a function that begins and ends at zero at its anchors in the continuum on opposite sides of the line profile. This feature can help distinguish a true signal from noise, as anything that produces a net area under the $V$ curve would be noise under these conditions. However, it should be noted that Stokes $V$ asymmetries originating with velocity gradients along the line of sight have been detected in solar magnetic fields \citep{1989A&A...221..338G}. Similar asymmetries, which may have the same origin, have been detected in the ultraweak fields of Am stars \citep{2011A&A...532L..13P,2016A&A...586A..97B,2018CoSka..48...53F}. Whether such asymmetrical Stokes $V$ signatures should be expected in hot star magnetospheres is not known. 

As for the true Stokes $V$ signal, two physical effects combine to produce it \citep{Gayley2015}. The first is due to the wavelength derivative of the Stokes $I$ profile itself, an effect that would be  present even if the line-of-sight B field were constant across the profile. The other is due to the gradient in the average line-of-sight B field across the profile, an effect that would be present even if the Stokes $I$ profile were constant (i.e., flat). Which of these dominates the signal at any wavelength can be established by the gradient in Stokes $I$: the first effect will only be dominant when the Stokes $I$ signal is strong, whereas deviations of the Stokes $V$ signal from the derivative of Stokes $I$ would clearly indicate that the second effect is active as well. The latter possibility shows that Stokes $V$ is not purely a diagnostic of the strength of the field, it is a combined diagnostic of field strength and field structure, requiring forward modelling to interpret.

%%%%%%%%%%%%%%%%%%%%%%%%%

\section{Circumstellar magnetic fields using the Hanle effect in resonance lines}\label{sec:hanle}

The Zeeman effect is an incoherent
process, where light polarization is manifested purely because of the 
energy splitting of the sublevels due to the action of a magnetic field. 
By contrast the Hanle effect involves linear polarization from resonance scattering 
arising from radiation anisotropy \citep[i.e., scattering polarization][]{1994ASSL..189.....S,2004ASSL..307.....L}.
The presence of a (weak) magnetic field modifies the distribution
of scattered light, altering the polarization from the non-magnetic case.
The Hanle effect operates mainly in a regime where atomic sublevels are only 
marginally separated -- at the level of the natural broadening
\citep{2002ApJ...568.1056C}.
The sublevels interfere quantum-mechanically with
each other, leading to phase coherence effects that govern the polarization and redistribution of polarized light, for example, resulting in changing polarization amplitude and rotation of the polarization position angle \citep{2008pps..book..247C}. 
Of chief importance for observations with {\em Polstar} is that the
Hanle effect applies to resonance line scattering, and many resonance
lines of hot massive stars are located in the UV waveband.  

Physically, the Hanle effect is relevant roughly when the Larmor frequency $\omega_B$ of the 
field is comparable to the Einstein A-coefficient of the transition, and can be characterized by a ``critical'' field value, $B_
H$. As an example for a
two-level atom $(J_0,J)$, $B_H$ is determined using
\begin{equation} \label{eq:Hanle_formula}
g_J\,\omega_B \sim A_{JJ_0}\;,\qquad \omega_B=(\mu_0/\hbar)\,B_H\;,
\end{equation}
where $g_J$ is the Land\'e factor of the upper level $J$, and $\mu_0$ is
Bohr's magneton.  Several relevant UV resonance lines accessible to \emph{Polstar} are given in Table~\ref{hantable}, showing that the mission can access a broad range of field sensitivity via the Hanle effect, roughly in the range of 1--100\,G.  

\begin{table}
\begin{center}
\caption{Sample FUV Lines for Hanle
Effect.  \label{hantable}} 
\begin{tabular}{cccccc}
\hline\hline Ion & $\lambda$ & $A$ & $g_u$ &
$g_{\rm eff}$ & $B_H$ \\
    &   (nm)    & $10^8$ s$^{-1}$ &   &   &  (G) \\ \hline
\ion{H}{i}      &121.57        &6.26   &1.33 &1      &53.5 \\ 
\ion{He}{ii} &164.03        &3.59   &1.33 &1      &30.7 \\ 
\ion{C}{ii} &133.45 &2.41   &0.8    &0.83 &34.3 \\ 
\ion{O}{iv}\,(1) &133.18 &2.58   &0.8    &0.83 &34.3 \\
\hphantom{\ion{O}{iv}}\,(2)    &134.30 &0.43   &0.8    &1.07  &6.1 \\
\hphantom{\ion{O}{iv}}\,(3)    &134.35 &2.57   &1.2    &1.10  &24.4 \\
\ion{C}{iv}     &154.82        &2.65 &1.33 &1.17 &22.6 \\ 
\ion{N}{v}      &123.88        &3.40   &1.33 &1.17 &29.0 \\ 
\ion{Mg}{ii}    &123.99        &0.0135 &1.33 &1.17 &0.115 \\ 
\ion{Si}{ii}    &126.04        &25.7   &0.8    &0.83 &365 \\
\ion{Si}{ii}    &180.80        &0.0254 &0.8    &0.83 &0.361 \\ 
\ion{Si}{iv} &139.38        &8.80   &1.33 &1.17 &75.1 \\ 
\ion{S}{ii}     &125.38 &0.512  &1.73 &1.87 &3.36 \\
\ion{S}{ii}     &125.95        &0.510 &1.6    &1.3    &3.63 \\ \hline \end{tabular}
\end{center}
\end{table}

The Hanle effect has been employed in solar research for decades, as a powerful diagnostic of weakly magnetized regions of the photosphere and chromosphere \citep{1994A&A...285..655F,1998A&A...337..565B,2004Natur.430..326T,2011ApJ...743...12M,
2016ApJ...830L..24D,2018ApJ...863..164D}, as well as of solar structures such as prominences and filaments, where radiation processes are dominated by scattering \citep{1977A&A....60...79L,1994SoPh..154..231B,2002Natur.415..403T,2003ApJ...598L..67C,2014A&A...566A..46O,2015ApJ...802....3M,2021A&A...647A..60B}. 
The Hanle effect in the FUV (e.g., \ion{H}{i} Lyman $\alpha$, \ion{O}{vi} 103.2\,nm doublet) has also been of particular interest for the plasma and magnetic diagnostics of the solar corona and wind \citep{1982SoPh...78..157B,1991OptEn..30.1161F,2002A&A...396.1019R,2011A&A...529A..12K,2019ApJ...883...55Z,2021ApJ...912..141Z}.

Applications to stars other than the Sun have been little addressed.
\cite{2011A&A...527A.120L}, \cite{2011A&A...530A..82I}, \cite{2012A&A...539A.122B}, and \cite{2012ApJ...760....7M} have
explored the influence of the Hanle effect in spectropolarimetry
of the photospheric lines of unresolved stellar atmospheres.  In
particular, \cite{2012ApJ...760....7M} calculated the total line polarization in Stokes
$Q$ and $U$ for stars with either dipole or quadrupole magnetic
fields.  Depending on field topology, strength, obliquity, and viewing inclination, those authors evaluated variations of the line-integrated polarization with rotational phase.  They found polarization amplitudes at the level of several tenths of a percent, depending on specific combination of parameters.

\begin{figure}
\centering
\includegraphics[width=\columnwidth]{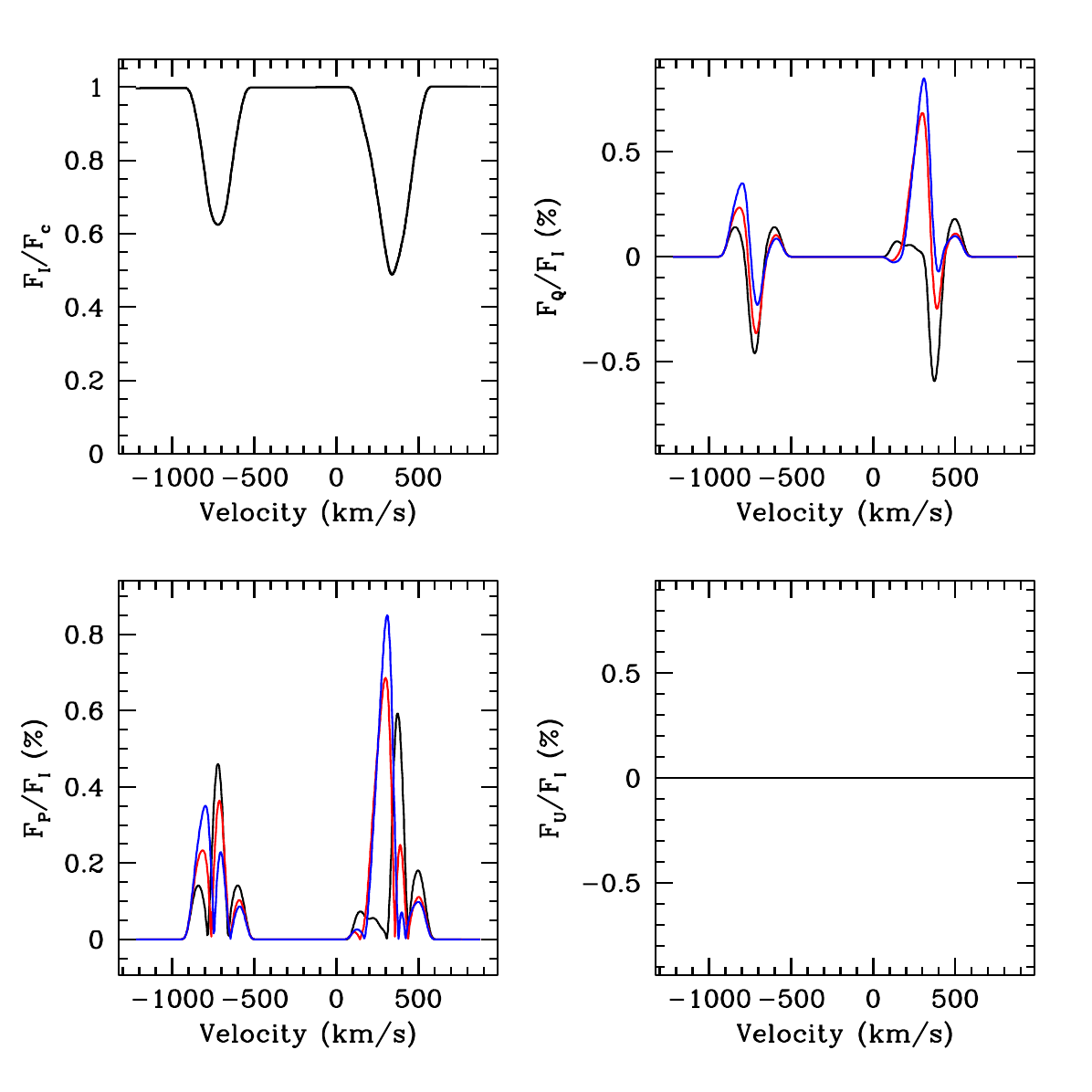}
\caption{Resolved polarization for rotationally broadened photospheric lines of the O~{\sc iv} triplet 133.9, 134.3, 134.4 nm (the latter two components are blended).  Results are shown in velocity shift relative to a mean wavelength of 134.2~nm for the triplet, weighted by the statistical weights. The panels are continuum normalized line profiles (upper left), polarized profiles as percent (lower left), Stokes $Q$ profiles (upper right), and Stokes $U$ profiles (lower right).  The rotating star is seen equator-on.  The dipole field is likewise equator-on (i.e., 90 deg from the dipole axis).  The black curve is for no magnetic field;  red is for $B_\ast=15$ G; blue is for 30~G. For an equator-on dipole, Stokes $U$ is unpolarized.}
\label{hanle1a}
\end{figure}

\begin{figure}
\centering
\includegraphics[width=\columnwidth]{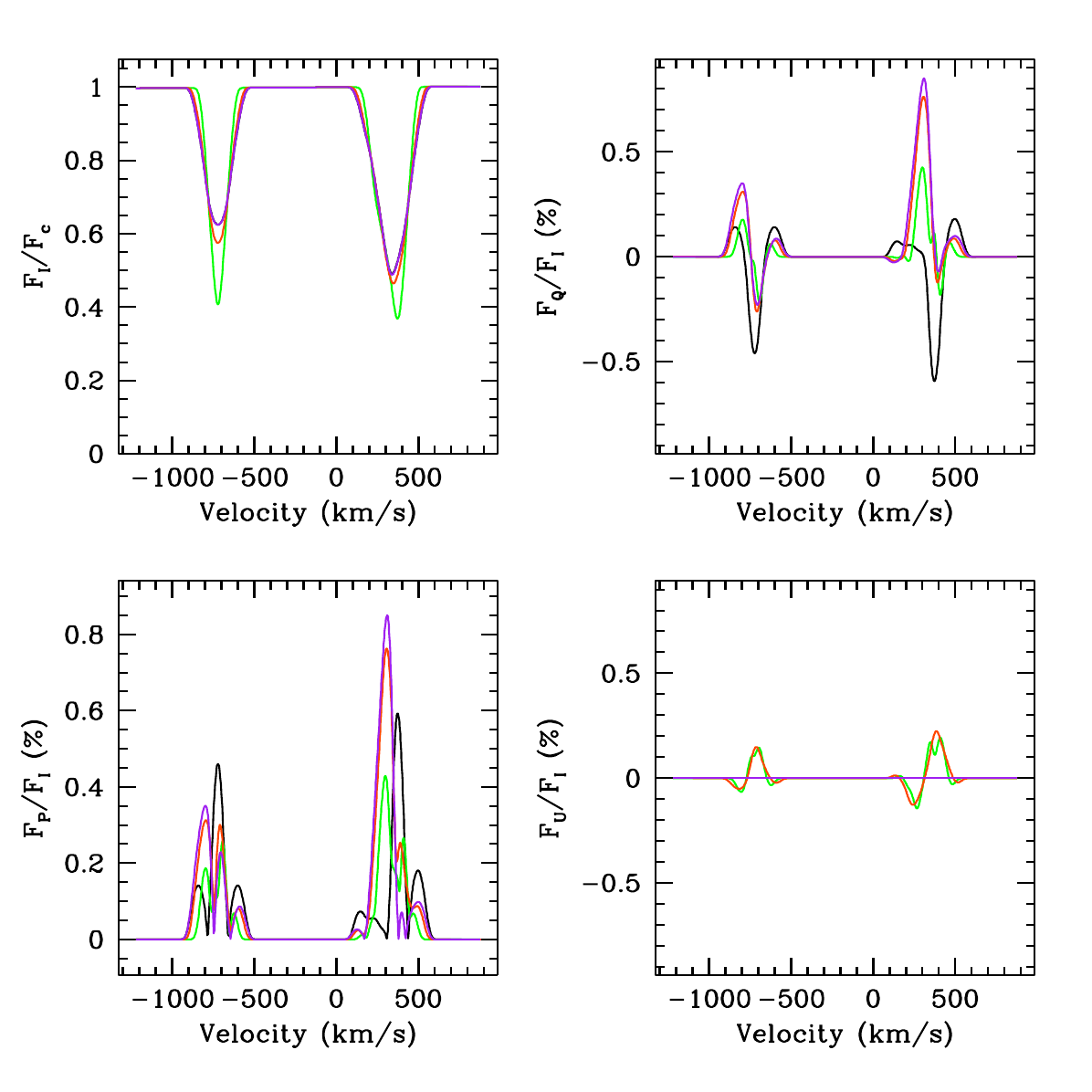}
\caption{Similar to Fig.~\ref{hanle1a}, now for a 30~G surface field with rotation axis inclinations of 30 (green), 60 (orange), and 90 (purple) degrees, keeping the magnetic axis aligned with the rotation axis.  An inclination of 0 deg thus corresponds to a view along the dipole field axis.  Black is the zero field case.}
\label{hanle1b}
\end{figure}

We extend the results of \cite{2012ApJ...760....7M} here in terms of resolved polarized line profiles specifically for a dipole magnetic field.
Figures~\ref{hanle1a} and \ref{hanle1b} show computational results for polarization in  rotationally broadened line profiles.  These illustrative examples are for the \ion{O}{iv} triplet 133.9, 134.3, 134.4 (components 1--3 in Tab.~\ref{hantable}), the latter two components being well blended.  The star is viewed equator-on with a rotation speed of 120 km/s, 
with the equator aligned along the negative Stokes $Q$ axis and the rotation axis along the positive $Q$ axis. 
Stokes $I$, $Q$, and $U$ line profiles are shown, either as continuum normalized for $I$ or relative polarization.
For Figure~\ref{hanle1a} the dipole field is seen equator-on (i.e., zero obliquity, with the magnetic axis aligned with rotational axis in the plane of the sky and the positive $Q$ axis).  Black is for zero magnetic field; red is for a polar surface field of 15~G (the ``Hanle'' field strength for this triplet); and blue is for 30~G.

Note that even without a magnetic field, the profile is polarized.  However, the line-integrated polarization is zero for $B_\ast = 0$.  Also note that Stokes $U$ polarization is zero for this scenario due to the symmetry of this geometry.  The Hanle effect leads to substantial alterations in the Stokes $Q$ profile by several tenths of a percent, as well as modifications to the polarization position angle across the profile.  Position angle rotation is signified by a change of sign in Stokes $Q$ for this figure.  The velocity shifts for the rotation are unchanged for the blue component, but the more blended red component leads to more complex behavior in the polarization.

Figure~\ref{hanle1b} is similar to Figure~\ref{hanle1a}, but now the field is always 30~G and  curves are for different
inclinations of the rotation axis at 30 (green), 60 (orange), and 90 (purple) degrees, keeping the dipole and rotation axes aligned,
with black still being the zero field case.  
When the field is viewed neither equator-on nor pole-on, a generally anti-symmetric Stokes $U$ profile results.  While the amplitude of polarization in $U$ is generally smaller than in $Q$, the differential changes owing to the Hanle effect are comparable.  Additionally, there are also relative changes in the Stokes $I$ line profiles themselves.  In one case ($i=30$; green), the Hanle effect changes the line depths to become nearly equal, whereas they are distinctly unequal in the zero field case.

The Hanle effect has also been explored in the context of circumstellar
media. Even without magnetism or the Hanle effect, resonance line scattering
can produce line polarization from scattering in circumstellar media and alter the shaping of wind emission profiles
\citep[e.g.,][]{1998A&A...332..686I, 1998A&A...337..819I, 2000A&A...363.1106I}.  Inclusion of the Hanle effect alters the line polarization in ways that depend on the strength and distribution of the magnetic field threading the circumstellar region.
\cite{1997ApJ...486..550I} and \cite{1999ApJ...520..335I}
used simplifying assumptions to investigate the
potential of the Hanle effect for tracing the magnetic field in
stellar winds.  Using a last scattering approximation combined
with the concept of the Sobolev optical depth, \cite{2004ApJ...609.1018I}
considered dipole field topologies and the ``WCFields'' model \citep{1998ApJ...505..910I} for
rotationally distorted winds \citep[wind compressed zone model,][]{1996ApJ...459..671I}.

\begin{figure}
\centering
\includegraphics[width=\columnwidth]{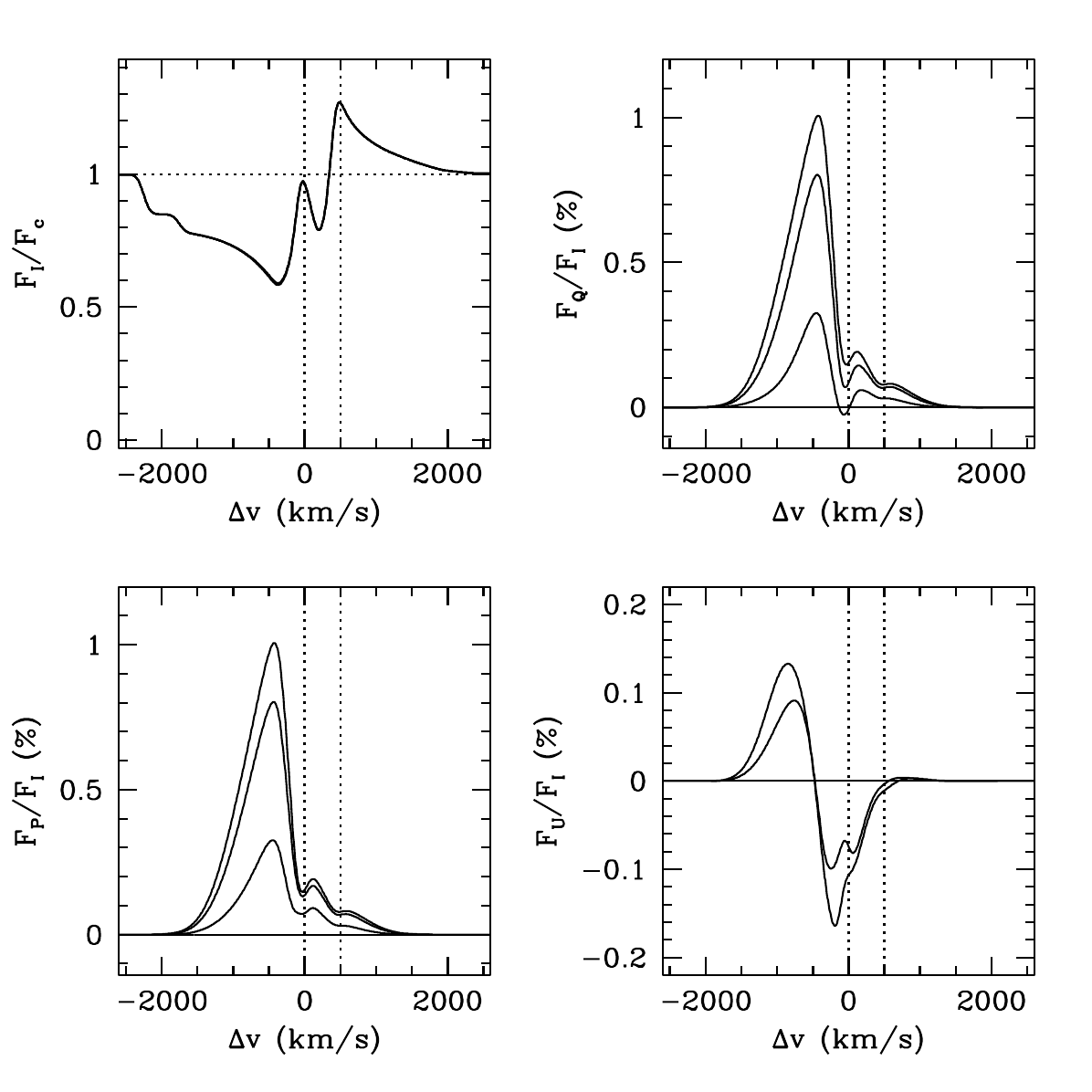}
\caption{The Hanle effect in the UV C{\sc iv} doublet for a spherical wind threaded by a dipole magnetic field.  The four panels follow the scheme
of Fig.~\ref{hanle1a}, with velocity shifts relative to the blue 154.8 nm component of the doublet.  The two vertical dotted lines are the rest wavelengths of the doublet components.  Four models were computed for a surface polar field of 100~G (note that no photosphere is included to focus on the wind effects).  For Stokes $I$, all four profiles overlap.  For Stokes $Q$ at upper right, there are 4 lines for obliquities of 0, 30, 60, and 90 degrees to the line of sight.  For pole-on there is no polarization across the profile.  The polarization becomes stronger as the dipole is viewed more side-on.  The same is true for the total polarization at lower left.  For Stokes $U$ both the pole-on and side-on dipole yield no net $U$ signal throughout the profile.}
\label{hanle3}
\end{figure}

Figure~\ref{hanle3} shows an example of
a  radiative transfer calculation for the Hanle effect in
the C{\sc iv} 155\,nm doublet.  The profiles were calculated using the HanleCLE code, which was derived from the HanleRT code of \cite{2016ApJ...830L..24D} in order to model stellar envelopes and coronae in the last-scattering approximation.
The two doublet components are indicated
by the vertical lines for their rest wavelengths in relative velocity.
The calculation is for an O~star at 30,000~K with wind terminal speed 2450 km/s (like $\zeta$ Pup)
and a typical $\beta=1$ wind velocity law.
For purposes of illustration, the wind is treated as spherically symmetric to highlight the Hanle effect with a dipole field superimposed.  The polar field strength is 90~G at the stellar surface.  Four cases are computed at field tilts of 0 (pole-on), 30, 60, and 90 (equator-on) degrees (again with the projection of the magnetic axis on the sky aligned with the positive $Q$ axis).  Note that the polarization is produced entirely by the shorter wavelength component of the doublet; the longer wavelength component produces no scattering polarization.  Best-case polarization levels at around 1\% over limited velocity shifts are easily measurable; overall, polarization levels at several tenths of a percent are produced at modest surface field strengths around $10^2$ G.

Figure~\ref{hanle3} highlights ways in which the
Hanle-effect diagnostic enabled by {\em Polstar} may be used to map the magnetospheres of massive stars.  There are 3
main considerations: (1) the Hanle effect is typically
sensitive to weaker magnetic fields when compared to the Zeeman
effect, (2) for the commonly strong Li-like resonance
doublets observed from massive star winds, one of the line components is insensitive
to the Hanle effect (unpolarizable upper level), and (3) the Hanle 
effect has different responses for different lines, so a multi-line 
approach is beneficial. 

Regarding item \#1, the Zeeman effect is certainly sensitive to
any field strength. However, its detectability depends in part 
on how the Zeeman splitting compares to the line broadening,
whether thermal, rotational, or from wind velocity distribution
\citep[e.g.,][]{Gayley2017}.  For many of the massive stars that would be targeted by {\em Polstar}, the Zeeman effect can still be employed to infer
the magnetism at the photosphere, and in limited cases
the circumstellar field near the photosphere. But these
sources will often have kG-level fields.  Thus the Hanle effect
will be employed for stars with lower surface fields, or as a tool for mapping the magnetic fields relatively far from the photosphere, in the wind acceleration zone or in the wind-confining magnetospheric lobes.

Regarding item \#2, the longer wavelength component of Li-like resonance doublets (upper level $J=1/2$) scatters isotropically to produce no polarization,
and therefore no Hanle effect. The shorter wavelength component 
(upper level $J=3/2$) scatters with a 50\% dipole-like efficiency 
(i.e., half dipole-like and half isotropic).
These effects are seen in Figure~\ref{hanle3}, where in Stokes-$I$, a wind line
is seen at each of the doublet components, but only one displays 
polarization.
The diagnostic value of this situation is that both of the doublet
components of C{\sc iv} form in the same spatial zones (one having half
the optical depth of the other). Consequently, any polarization
at the wavelength of the non-polarized component must arise from
the source continuum polarization or from interstellar polarization,
or both.  Such a polarization serves as an internal
calibration that can be
applied to the other doublet component to infer the polarization
arising strictly from line resonance scattering and the Hanle effct.

Regarding item \#3, it is difficult to know {\em a priori}
whether the resonance line polarization is influenced by the
Hanle effect, because of the need to know first the polarization 
the line would produce in the absence of a magnetic field. The
fact that the interpretation of the Hanle effect relies on the
modeling of such a zero-field polarization is arguably the main 
limitation of this diagnostic. On the other hand, there
are multiple viable resonance lines (c.f., Tab.~\ref{hantable})
available for analysis, each with different Hanle sensitivites (i.e., different characteristic $B_H$ values).
As \cite{1997ApJ...486..550I} have discussed, it is possible to disentangle the 
Hanle effect from non-magnetic resonance scattering polarization 
through analysis of the differing responses of the different lines. Such a ``differential'' Hanle effect is also routinely adopted as a diagnostic of solar magnetism \citep[e.g.,][]{Stenfloetal1998}.
In this way, the 3D magnetic field in the region of line formation 
can be reconstructed through simultaneous model fitting of multiple lines.

Not only will {\em Polstar} provide exquisite
spectropolarimetric capability at UV wavelengths, but the mission
emphasizes the importance of time-series observations.  Since direct
imaging is not an option for these distant stars, the only viable path
for mapping the 3D distribution of magnetic, density,
and velocity structures about the stars is by seeing the system
from varying perspectives.  This naturally arises through stellar
rotation, but requires numerous visit to the same targets.  It is
this principle that will be employed for extracting a picture of
circumstellar structure from the spectropolarimetric data at different
rotational phases.  In the case of the Hanle effect, this makes use
of the high levels of line broadening, well beyond thermal broadening,
owing to rotation ($\sim 10^2$ km/s) and/or wind flow ($\sim 10^3$
km/s).

\begin{table}
\begin{center}
\caption{Targets for Hanle Studies.  \label{hantargs}}
\begin{tabular}{ccccc}
\hline\hline Name &	$\eta_\ast$    & $v_\infty$ & $v_{\rm rot}$ & Rationale \\
     &          & (km/s)  & (km/s)  & \\ \hline
$\beta$ CMa        &13.5    &1885    &24      &good confin. \\
$\epsilon$  CMa        &14.5    &1720    &21      &good confin. \\
$\zeta$ Ori        &0.20    &2000    &127+    &weak confin. \\
$\tau$  Sco        &7.9     &2540    &8       &complex field \\
$\zeta$ Cas        &430     &1880    &58      &strong confin. \\
HD 64740        &2510    &---     &---     &strong confin. \\ \hline
\end{tabular}
\end{center}
\end{table}

{\em Polstar} could for the first time allow for routine measurement
and diagnostic use of the Hanle effect in other stars, representing
a fruitful new collaboration between heliophysics and stellar astrophysics.
For use of the Hanle effect for massive star studies, we expect 
initially a limited selection of targets. 
Table~\ref{hantargs} provides a preliminary list of targets in which to
investigate the Hanle effect in photospheric and circumstellar
spectral lines.
This selection spans a range of wind magnetic confinement parameter ($\eta_\ast$) values \citep[see][]{ShultzTC} and field topologies. The table also indicates the wind terminal speed
and the stellar equatorial rotation speed.
Based on the last column of the table, our
selection of targets would allow {\em Polstar} to probe circumstellar magnetism ranging from wind-dominated cases ($\eta_\ast\ll 1$) to wind-confined scenarios and dynamical magnetospheres ($\eta_\ast \sim 10$) to highly confined and large centrifugal magnetospheres ($\eta_\ast
\gg 1$).  Not only can we measure circumstellar winds, but will
also test model predictions for the Hanle effect in photospheric
lines.

\section{Summary and Conclusion}\label{Conclusion}

We have investigated the possibility of detecting and characterizing magnetic fields using UV spectropolarimetry, both in stellar photospheres and magnetospheres. 
This can provide qualitatively new information about magnetic fields in winds through both the Zeeman and Hanle effects, complemented with simultaneous photospheric magnetic field measurements.

Considering magnetic fields in the photospheres of hot stars detected through the Zeeman effect, we find the LSD technique should be highly efficient, and that magnetic fields should be detectable at achievable S/N for instruments currently being planned.  We used synthetic spectra with varying synthetic noise to assess the detectability of photospheric magnetic fields and the efficiency of LSD in the UV.  
Due to the multi-line nature of LSD, the increased density of lines in the UV compared to the visible enhances the efficiency of this method.  This more than compensates for signals being weaker in individual lines due to the wavelength dependence of the Zeeman effect, making magnetic fields of hot stars more detectable in the UV.  
For the parameters of the proposed {\em Polstar} mission, we find magnetic field uncertainties as low as 50 G may be reached for a S/N of 200 in B stars with a moderate \vsini~$=30$ \kms\ (even weaker for lower \vsini), implying magnetic fields of $\sim$150 G could be routinely detectable at a S/N of $\sim$200.  These results also suggest that, at a similar S/N, {\em Polstar} would be able to detect magnetic fields a factor of $\sim$2 weaker than the current best high resolution spectropolarimeters in the visible, for O and B stars.  
In practice, the smaller collecting area of a spaced based UV mission will likely offset the gain in stellar flux and improved sensitivity from LSD, relative to ground based visible observations, but detecting photospheric magnetic fields should be possible with reasonable exposure times.
In addition to detecting magnetic fields, the same sensitivity and high resolution will allow {\em Polstar} to characterize the geometry of these fields using time series observations and tomographic methods like ZDI. 

The magnetic fields around hot stars may also be detectable by using the Zeeman effect in wind-sensitive UV resonance lines. We have extended the UV-ADM model initially reported by \citet{erb21} to model magnetospheric polarization, from which we have produced synthetic polarized (Stokes~$V$) line profiles. While these results are preliminary, they suggest that a polarization signal would be detectable at a few tenths of a percent. Observing the rotational modulation of these polarization signals would allow for the reconstruction of the magnetic field distribution in the magnetosphere.

The Hanle effect is complementary to Zeeman diagnostics.  Observations of the Hanle effect rely on resonance lines, which for hot stars are almost entirely in the UV.  In semi-classical terms, the Hanle effect relates to Larmor precession of atomic oscillators during a resonance line scattering.  When the Larmor frequency is comparable to the radiative rate, the distribution of scattered light and its polarization with direction is modified from the case of no magnetic field.  Thus an analysis of multiple lines can be used to reconstruct the properties of photospheric and/or circumstellar magnetism.  
The Hanle effect and the Zeeman effect are both manifestations of an interaction between a photon and an atom in the presence of a magnetic field.
The distinction becomes useful in typical applications, such as the longitudinal Zeeman effect for weak fields producing circular polarization and the Hanle effect for resonance line scattering producing linear polarization. 
In practice, the Hanle effect tends to be sensitive to modest magnetic field strengths of 10--100G, and in ultraviolet lines of hot massive stars, is expected to produce polarizations at the level of tenths of a percent that would be measurable with a space-borne facility like {\em Polstar}.

These results highlight the potential for UV spectropolarimetry, at high resolution with wide wavelength coverage, to be a powerful new tool for observing and characterizing magnetic fields and stellar winds.  The {\em Polstar} mission is excellently suited to provide these observations.  This will provide the ability to simultaneously characterize photospheric and circumstellar magnetic fields, over a wide range of field strengths.  That will in turn inform us about photospheric processes, wind processes, and how they interconnect to control angular momentum loss, mass loss, and how they impact both stars and their surrounding environments.

\backmatter

\bibliography{bibliography}

\section*{Statements \& Declarations}

\subsection*{Funding}
C.E. gratefully acknowledges support for this work provided by NASA through grant number \textbf{HST-AR-15794.001-A} from the Space Telescope Science Institute, which is operated by AURA, Inc., under NASA contract NAS 5-26555.

R.I. and C.E. gratefully acknowledge that this material is based upon work supported by the National Science Foundation under Grant No. AST-2009412.

M.E.S. acknowledges financial support from the Annie Jump Cannon Fellowship, supported by the University of Delaware and endowed by the Mount Cuba Astronomical Observatory.

G.A.W. acknowledges Discovery Grant support from the Natural Sciences and Engineering Research Council of Canada (NSERC).

\subsection*{Author Contribution}
All authors contributed to the preparation of the manuscript.  The modelling and analysis in Sect.\ 2 was primarily done by K.H., C.P.F., and G.A.W.  The analysis and modelling in Sect.\ 3 was primarily completed by C.E., with advisement from V.P. The analysis and modelling in Sect.\ 4 was primarily done by R.I., R.C., T.P.A., and R.M.S.
 
\subsection*{Competing Interests}
The authors have no relevant financial or non-financial interests to disclose.  

\subsection*{Data Availability}

The UV-ADM code and model grid used for this work is available from author C. Erba upon request.  The Least Squares Deconvolution code used is available at https://github.com/folsomcp/LSDpy.  Other data used are available from the corresponding author on reasonable request.

%\begin{appendices}
%\section{Section title of first appendix}\label{secA1}
%\end{appendices}

%\newpage
%\onecolumn
%\centering
\section*{Affiliations}
\noindent
$^{1}$Tartu Observatory, University of Tartu, Observatooriumi 1,T\~{o}ravere, 61602, Estonia 
\smallskip

\noindent
$^{2}${\orgdiv{Department of Physics and Astronomy}, \orgname{East Tennessee State University}, \orgaddress{
\city{Johnson City}, \postcode{37614}, \state{TN}, \country{USA}}}
\smallskip

\noindent
$^{3}${\orgdiv{High Altitude Observatory}, \orgname{National Center for Atmospheric Research}, \orgaddress{\street{P.O. Box 3000}, \city{Boulder}, \state{CO}, \postcode{80307-3000}, \country{USA}}}
\smallskip

\noindent
$^{4}${\orgname{Instituto de Astrof\'isica de Canarias}, \orgaddress{\postcode{E-38205} \city{La Laguna}, \state{Tenerife}, \country{Spain}}}
\smallskip

\noindent
$^{5}${\orgdiv{Departamento de Astrof\'isica}, \orgname{Universidad de La Laguna}, \orgaddress{\postcode{E-38206} \city{La Laguna}, \state{Tenerife}, \country{Spain}}}
\smallskip

\noindent
$^{6}${\orgdiv{Department of Physics \& Astronomy}, \orgname{University of Iowa}, \orgaddress{\street{203 Van Allen Hall}, \city{Iowa City}, \postcode{52242}, \state{IA}, \country{USA}}}
\smallskip

\noindent
$^{7}${\orgdiv{Department of Physics, Engineering Physics and Astronomy}, \orgname{Queen’s University}, \orgaddress{\city{Kingston}, \state{ON}, \country{Canada}, \postcode{K7L 3N6}}}
\smallskip

\noindent
$^{8}${\orgdiv{Max Planck Institute for Solar System Research}, \orgname{Justus-von-Liebig-Weg 3}, \orgaddress{\city{G\"{o}ttingen}, \country{Germany}, \postcode{37077}}}
\smallskip

\noindent
$^{9}${\orgdiv{LESIA, Paris Observatory}, \orgname{PSL University, CNRS, Sorbonne Université, Université Paris Cit\'e}, \orgaddress{\street{5 place Jules Janssen}, 
\postcode{F-92195}, \state{Meudon}, \country{France}}}
\smallskip

\noindent
$^{10}${\orgdiv{Department of Physics and Astronomy, Bartol Research Institute}, \orgname{University of Delaware}, \orgaddress{
\city{Newark}, \postcode{19716}, \state{DE}, \country{USA}}}
\smallskip

\noindent
$^{11}${\orgdiv{Dept. of Physics \& Space Science}, \orgname{Royal Military College of Canada}, \orgaddress{\street{PO Box 17000, Station Forces}, \city{Kingston}, \state{ON}, \country{Canada}, \postcode{K7K 7B4}}}
\smallskip

\end{document}